\documentclass{ws-ijmpd}
\usepackage[super,compress]{cite}
\usepackage{color}
\usepackage{placeins}

\usepackage{amssymb}
\usepackage{amsmath}
\usepackage{amscd}
\numberwithin{equation}{section}

\usepackage{hyperref}
\usepackage{xcolor}
\hypersetup{
     colorlinks   = true,
     citecolor    = blue
}

\newcommand{\complex}		{\mathbb{C}}
\newcommand{\reals}			{\mathbb{R}}

\newcommand{\w}[1]			{\mathcal{W}{(#1)}}

\begin{document}


%
\catchline{}{}{}{}{}
%

\title{FLUID DYNAMICS IN THE ELLIS WORMHOLE
}

\author{G. BUTBAIA$^1$ and Z.N. OSMANOV$^{1,2}$
}

\address{$^1$School of Physics, Free University of Tbilisi\\
Tbilisi, 0159, Georgia\\
$^2$E. Kharadze Georgian National Astrophysical Observatory\\
Abastumani, 0301, Georgia.\\
z.osmanov@freeuni.edu.ge
}

\maketitle

\begin{history}
\received{Day Month Year}
\revised{Day Month Year}
\end{history}

\begin{abstract}
In the Ellis wormhole metrics we study characteristics of fluid dynamics and the properties of linear sound waves. By implying the energy-momentum equation and the continuity equation in the general relativistic manner we examine the flow dynamics and solve the corresponding equations for a relatively simple case - radial flow. To study the linear sound waves the equations governing the mentioned physical system are linearized and  solved and interesting characteristic properties are found.
\end{abstract}


\ccode{PACS numbers: Need to be added!}

\section{Introduction}

The existence of a certain class of solutions of the Einstein field equations called wormholes (WH) is still a matter of discussion. It is assumed that these WHs might be formed during the inflation epoch of the evolution of the universe [\refcite{wheeler}-\refcite{hawking}]. In general it is natural to assume that the central region of  WHs might be composed of accreting matter [\refcite{accret}]. The study of dynamics of matter inside the WH metrics might be a significant step to identify these exotic objects. 

The simplest WH was proposed by Ellis [\refcite{ellis}], where the author, by analysing the Einstein vacuum field equations, has obtained the solutions of a traversable WH, comprising identical three dimensional areas connected by a relatively narrow throat. According to the study, the Ellis WH is "transparent" to a particles moving inside it. Therefore, matter can flow from one distant end of the WH to the other [\refcite{Brk}].

In general it is believed that WHs can provide magnetic fields as strong as $10^{13}$G [\refcite{kns}], which means that if the charged particles are trapped by rotating magnetic field lines, dynamics might reveal interesting features. A single particle approach has been proposed in the work [\refcite{arsena}], where by implying the method developed in [\refcite{gud}, \refcite{rdo}] it has been shown that if the magnetic field lines are twisted and are lagged behind the rotation, in certain cases, particles can leave the WH region.

A single particle approximation should be generalised by an ensemble of particles, in which context the work presented in [\refcite{whshadExt}] is worth mentioning. In particular, the authors consider the Ellis WH and optically thin disk around it. Two types of solutions have been found: the static solution regardless of the density profile and the solution describing dust entering the WH metrics from one side and escaping from the other. Effectively the matter flowing inside the WH might generate interesting effects in the context of dark matter. In particular, in the work by Kirillov and Savelova [\refcite{dark}] density perturbations is considered and it is shown that at large distances WHs  behave like non-barionic particles reproducing the features of dark matter.

Generally speaking, the study of gas dynamics in the WH metric is a quite complex task, because the behaviour of matter should be described not only by relativistic effects but MHD approximation might be important since, as we have already mentioned, magnetic field might be present in WHs. Although, as a first step we consider the hydrodynamic flow and study the two subtasks: the time stationary flows and sound waves for the hot gas. For this purpose we will imply the relativistic Euler equation and the continuity equation respectively, expand the equations up to the first order terms, numerically solve them and analyse the obtained solutions.

The paper is organized in the following way: in Sec.~\ref{sec:drainhole} we derive the main equations governing the considered physical system in the Ellis-Bronnikov metrics, in Sec.~\ref{sec:discusion} we solve the derived equations for the time-stationary flows and consider the linear sound waves and in Sec.~\ref{sec:summary}.
we summarize our results.


\section{Main Consideration}\label{sec:drainhole}
In this paper we consider the Ellis-Bronnikov wormhole [\refcite{ellis}, \refcite{Brk}] which is a pseudo-Riemannian $4$-manifold $\w{a}$ ($a > 0$) having the following metric tensor:
	\begin{gather}
	ds^2 = g_{\mu\nu}dx^{\mu}dx^{\nu} = \\
		= -d{t}^2 + d{r}^2 + \left(a^2+r^2\right)\left(d{\theta}^2+\sin^2{\theta}~d{\phi}^2\right),\label{eq:metric}
	\end{gather}
where $r,t\in\reals$, $\theta\in(0,\pi)$ and $\phi\in (-\pi,\pi)$. Here $t$ is time, $r$ is a coordinate which measures the proper radial distance, the parameter $a$ is referred to as the throat radius of the WH and $\theta$ and $\phi$ are the spherical coordinates. Throughout the paper we use the unit $c = 1$, where $c$ is the speed of light. In general there is other class of WHs which are characterised by intrinsic gravity, but the aim of the present paper is to study flow dynamics and propagation of sound waves in the most simplest form of WH, because the problem itself is quite complex. Another approximation we use is that we look for solutions which do not depend on $\phi$ and $\theta$.

To derive the fluid equations we first introduce the energy momentum tensor and current:
\begin{gather}
T^{\mu\nu} = (p + e) u^\mu u^\nu + pg^{\mu\nu}~~\text{and}~~J^\mu = \rho u^\mu,\qquad\label{eq:tensors}
\end{gather}
where $u^{\mu} = dx^{\mu}/ds$ represents the four velocity, $\rho$ is density and
$e$ and $p$ are energy and pressure, respectively. It is straightforward to show that from the conservation laws of energy-momentum and mass [\refcite{rezzolla}]:
\begin{gather}\label{moment1}
\nabla_{\mu}T^{\mu\nu} = 0,
\end{gather}
\begin{gather}\label{cont1}
\nabla_{\mu}J^{\mu} = 0,
\end{gather}
one arrives at the following set of equations governing the physical system for purely radial motion:

	\begin{gather}
		\frac{\partial v}{\partial t} + v\frac{\partial v}{\partial r} = -\frac{1-v^2}{\rho}\left[\frac{\partial \rho}{\partial r} + v\frac{\partial \rho}{\partial t}\right]c_s^2, \label{momcont}\\
		\frac{\rho}{1-v^2}\left[\frac{\partial v}{\partial r} + v\frac{\partial v}{\partial t}\right]	 + \frac{2\rho r}{a^2+r^2}v = -\frac{\partial \rho}{\partial t} - v\frac{\partial \rho}{\partial r},\nonumber
	\end{gather}

where $v = dr/dt$ and we have taken into account that the temperature is ultra-relativistic, when $p = e/3$, $e = \kappa\rho^{4/3}$ and thus the speed of sound is given by $c_s = 1/\sqrt{3}$.

\section{Discusion}\label{sec:discusion}
For the most of the astrophysical objects it is observationally evident that the flows are stationary, therefore, in this section, by applying the aforementioned equations (see Eqs. (\ref{momcont})) we will numerically solve the equations in the time-stationary approximation. Another task we are going to consider is the propagation of sound waves in a static flow to study the corresponding properties in the liner regime.

\subsection{Time-stationary solutions}

Of particular interest are the solutions which are stationary in time, that is the terms $\partial_t v$ and $\partial_t\rho$ vanish. Thus reducing Eqs. (\ref{momcont}) to the following coupled ODE system:
	\begin{gather}
		\frac{\partial v}{\partial r} = -\frac{1-v^2}{1-v^2/c_s^2}\frac{2rv}{a^2+r^2}, \qquad
		\frac{\partial \ln{\rho}}{\partial r} = \frac{v^2}{c_s^2-v^2}\frac{2r}{a^2+r^2}.\label{eq:stationary}
	\end{gather}
	
	This allows us to construct an initial value problem (IVP) with the following initial conditions $v(0) = v_0$ and $\rho(0) = \rho_0$ for some reals $-1< v_0 < 1$ and $\rho_0>0$. Furthermore, we shall show that there exists a critical velocity $v_c = c_s= 1/\sqrt{3}$ and in order to ensure a stable solution centered at the throat of the wormhole, the initial velocity must satisfy the following condition $v_0 \neq v_c$.
	
	In order to solve the IVP, we integrate Eqs.~\eqref{eq:stationary} to find that the velocity is given by an implicit equation:
	\begin{gather}
		\ln\left(\frac{1-c_s^2}{2c_s^2}|v(1-v^2)|\right) = -\ln|a^2+r^2| + C_1,~\text{which for}~c_s=1/\sqrt{3}~\text{reduces to:}\nonumber \\
		v(r)\left[1-v(r)^2\right] = \frac{a^2}{a^2+r^2}v_0(1-v_0^2),\quad v(0)=v_0,\label{eq:velocitySolution}
	\end{gather}
whereas the density $\rho(r)$ satisfies the following equation:
	\begin{gather}
		\rho(r) = \rho_0\left|\frac{1-v^2(r)}{1-v_0^2}\right|^{1/2c_s^2}.\label{eq:densityStationary}
	\end{gather}
	Note that $\rho(r)$ is bounded and attains its maximum value when the velocity is zero.

\begin{figure}[h]
	\begin{center}$
		\begin{array}{c}
		\includegraphics[width=60 mm]{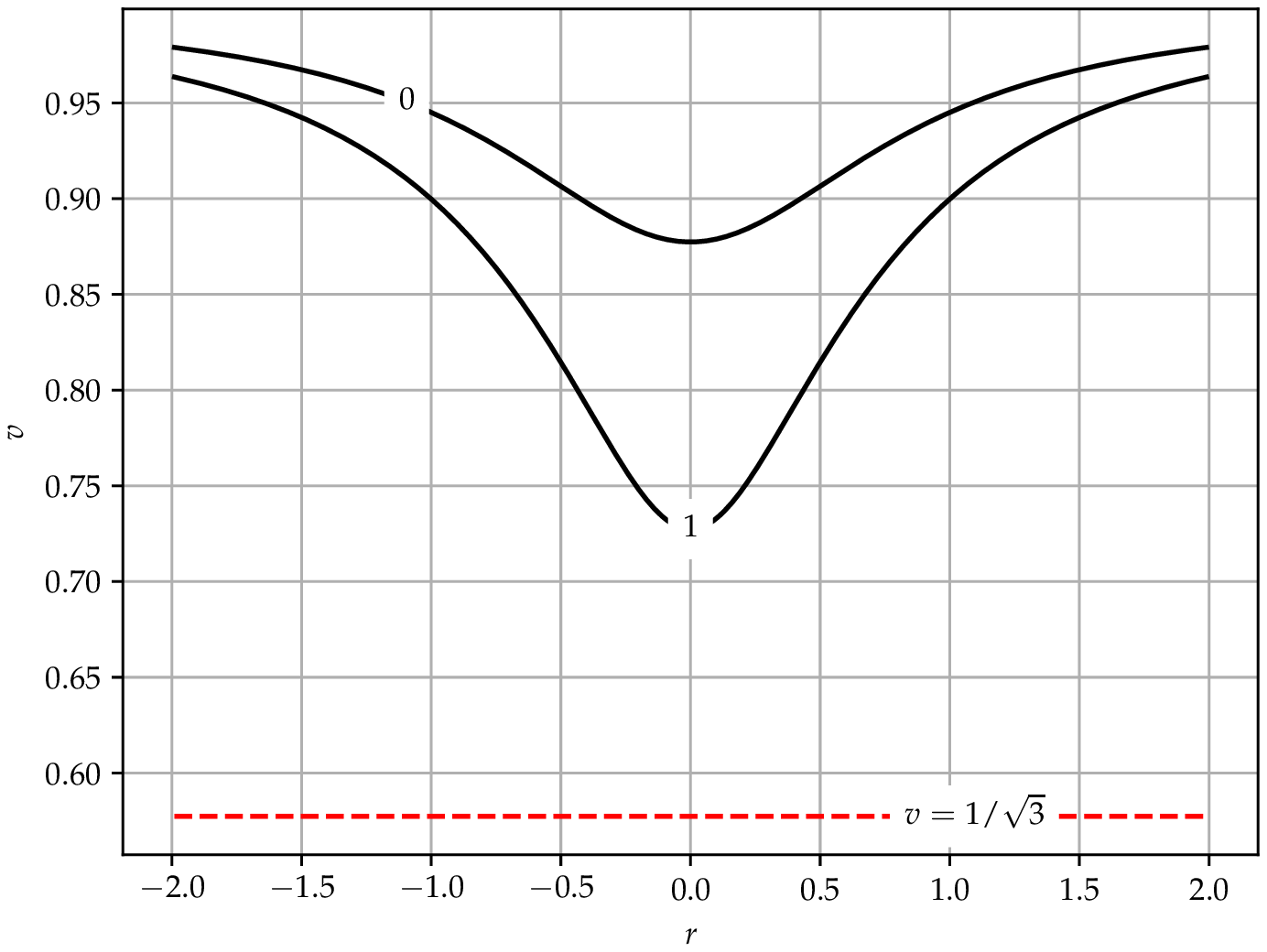}
		\includegraphics[width=60 mm]{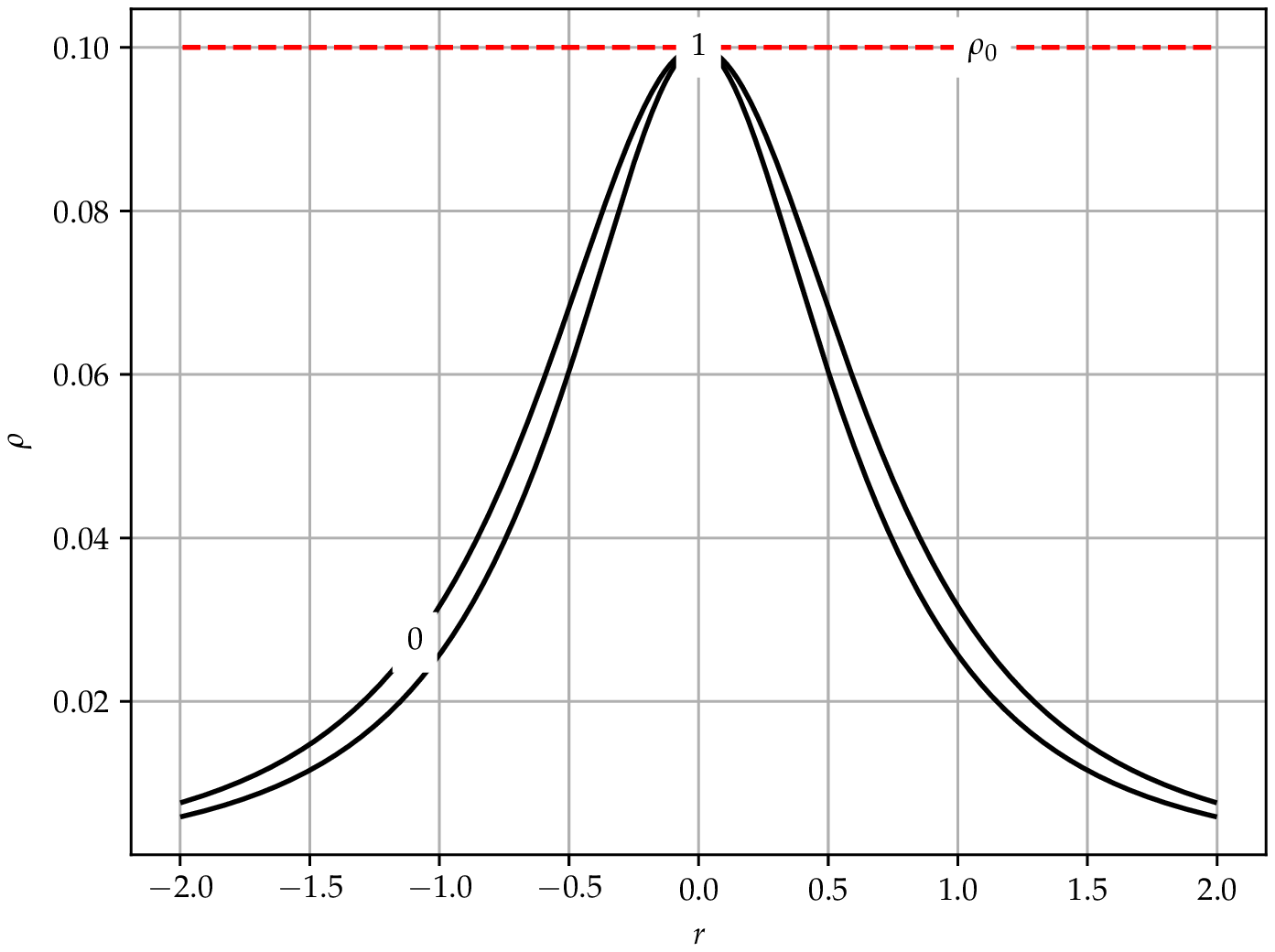} \\
		\includegraphics[width=60 mm]{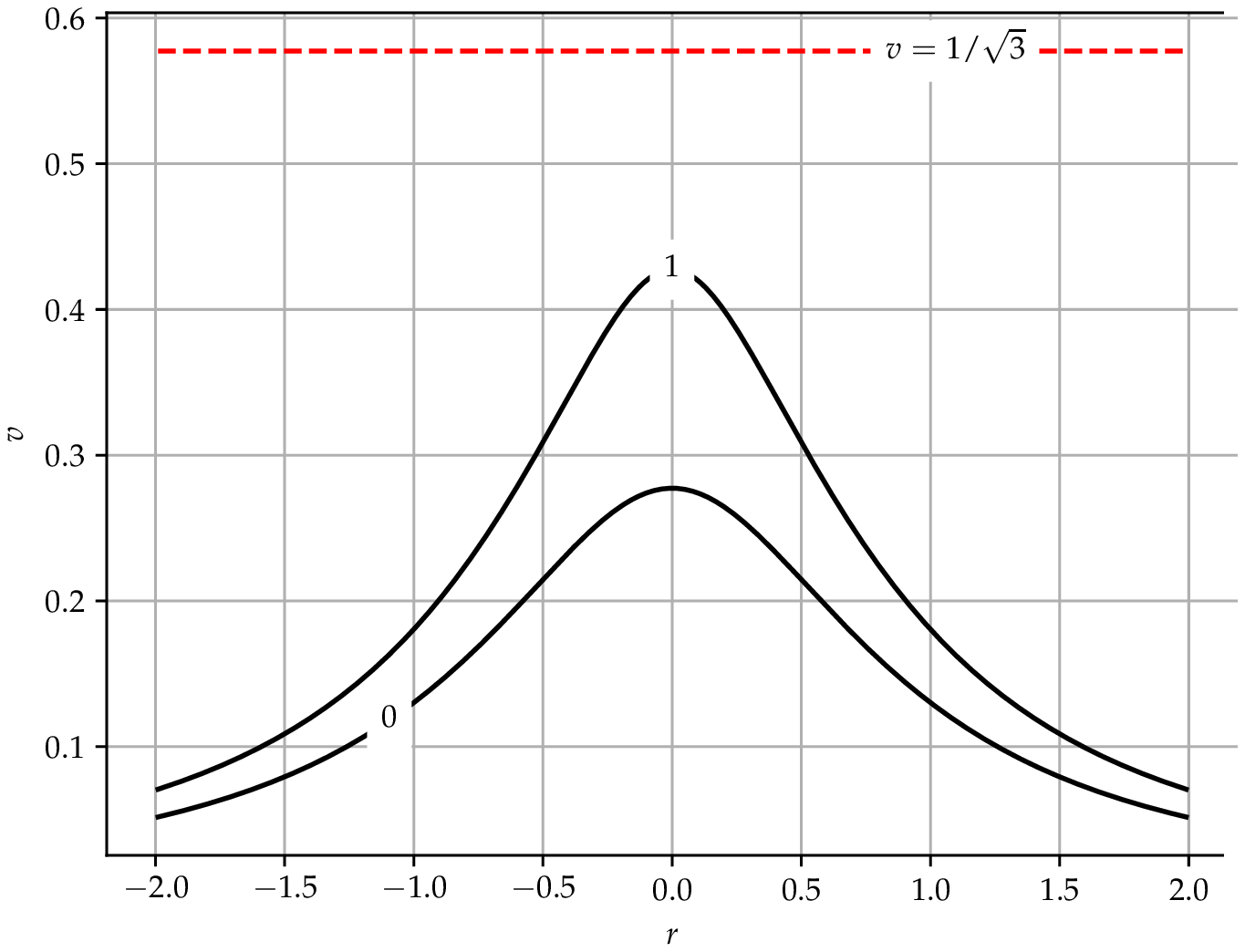}
		\includegraphics[width=60 mm]{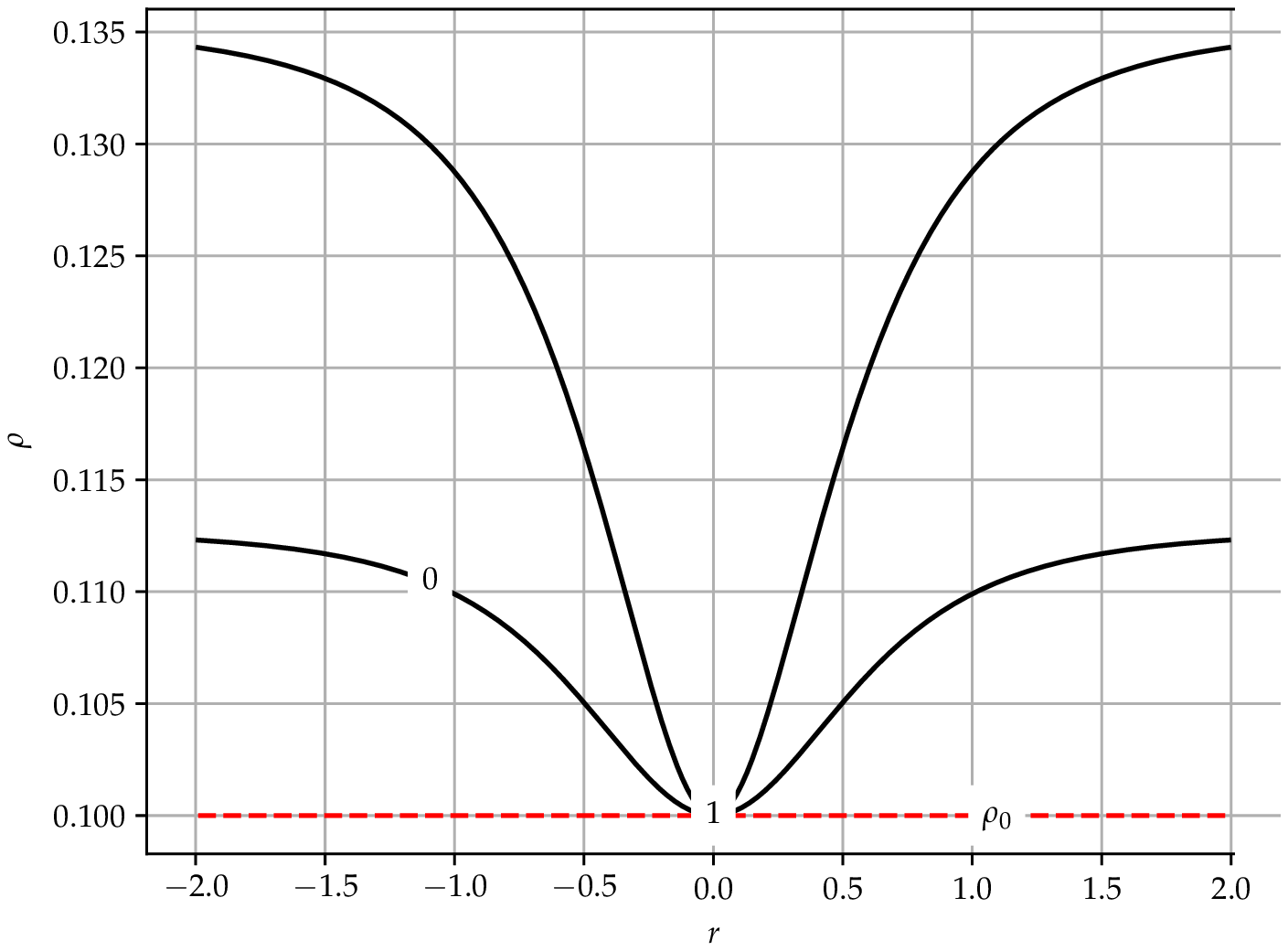}
		\end{array}$
	\end{center}
	\caption{A family of solutions of the Eq.~\eqref{eq:stationary} and Eq.~\eqref{eq:velocitySolution}. Top row shows solutions for $v(r)$ and $\rho(r)$ for $v_0 > v_c$ whereas the bottom row shows solutions for $v_0 < v_c$.}
	\label{fig:velrho}
	\end{figure}

	Several example for the numerical solutions of the $v(r)$ and $\rho(r)$ are shown on Fig.~\ref{fig:velrho}. As it is clear from the plots there are two different classes of solutions and thus two classes of stationary flows. One class of solutions correspond to the initial velocity $v_0>c_s$ (top row), when far from the central zone the velocity tends to the speed of light and consequently the density asymptotically vanishes reaching its maximum value in the centre of the WH. This can also be seen from Eq. \eqref{eq:densityStationary}.

Another class of solutions correspond to $v_0<c_s$ (bottom row) and unlike the previous case, far from the centre the velocity tends to zero and density asymptotically tends to a certain constant value. 
	
	A typical solution of Eqs.~\eqref{eq:stationary} involves either the maximum of velocity $v(r)$ or the maximum of density $\rho(r)$ centered at the center of the throat of the WH, but not both. This can be shown by observing second order derivatives of Eqs.~\eqref{eq:stationary} at $r=0$, which are given by:
	\begin{gather}
		\frac{\partial^2 v}{\partial r^2}\bigg\vert_{r=0} = -\frac{2c_s^2 v_0(1-v_0^2)}{a^2(c_s^2-v_0^2)},\qquad \frac{\partial^2 \rho}{\partial r^2}\bigg\vert_{r=0}	 = \frac{2\rho_0 v_0^2}{a^2(c_s^2-v_0^2)}.\label{eq:2ndorderderivatives}
	\end{gather}
	Hence, we see that $\mathrm{sgn}\left(\partial_r^2v\right)\vert_{r=0}=-\mathrm{sgn}\left(\partial_r^2\rho\right)\vert_{r=0}$, thus implying that one's maximum is the other's minimum.

\begin{figure}[h] 
	\begin{center}$
		\begin{array}{c}
		\includegraphics[width=70 mm]{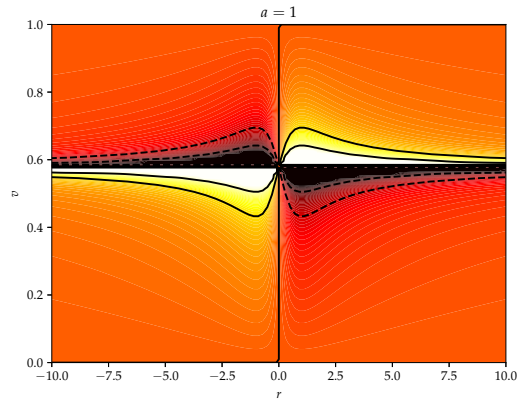}
		\end{array}$
	\end{center}
	\caption{A plot of the $\displaystyle f(r,v) = \frac{\partial v}{\partial r}$. Observe the clear separation line at $v_c = c_s = 1/\sqrt{3}$.}
	\label{fig:phase}
	\end{figure}

	This can also be seen by inspecting the contour-plot of the phase-space diagram at each radial point $r\in\reals$ for the velocity shown on Fig.~\ref{fig:phase}.

Since the solutions have no singularities it is clear that the flow can penetrate the WH from one side and leave it from the other. This in turn is a key property of any WH. In the light of a recent study, according to which the active galactic nuclei might be traversable WHs [\refcite{agn}], the obtained results might be quite prospective and intriguing.


\subsection{Sound waves}\label{sec:stability}
Generally speaking, a non-trivial geometry of the Ellis-Bronnikov WH might influence not only the flow dynamics but it can affect the physical characteristics of sound waves as well. For this purpose in this subsection we focus on analyzing the linear waves - sound waves. Our goal here is to show that given a constant solution $(\rho_0,~v_0 = 0)$, its perturbation is stable. We do this by defining $\epsilon$ as the perturbation parameter, and expanding the perturbed solution $(\tilde{\rho}_\epsilon,~\tilde{v}_\epsilon)$ as a formal power series:
	\begin{gather}
		\tilde{\rho}_\epsilon = \sum_{n=0}^\infty \epsilon^n \rho_n,\label{eq:perturb} \\
		\tilde{v}_\epsilon = \sum_{n=0}^\infty \epsilon^n v_n.\nonumber
	\end{gather}
	Substituting Eqs.~\eqref{eq:perturb} into the Eqs.(\ref{momcont}) yields the following first order coupled PDE system for the first order terms:
	\begin{gather}
		\rho_0\frac{\partial v_1}{\partial t} + c_s^2\frac{\partial \rho_1}{\partial r} = 0, \label{eq:linear} \\
		\rho_0\frac{\partial v_1}{\partial r} + \frac{\partial \rho_1}{\partial t} + \frac{\displaystyle 2\rho_0 r}{a^2+r^2}v_1 = 0,\nonumber
	\end{gather}
where $c_s^2 = 1/3$. This system can be reduced to the following hyperbolic PDE:
	\begin{gather}
		\frac{1}{a^2+r^2}\frac{\partial}{\partial r}\left[(a^2+r^2)\frac{\partial\rho_1}{\partial r}\right] = \frac{1}{c_s^2}\frac{\partial^2\rho_1}{\partial t^2}\label{eq:linDensity}
	\end{gather}
	It can be immediately observed that  Eq.~\eqref{eq:linDensity} reduces to the wave equation in spherical coordinates in the limit $a\rightarrow 0$.  		
	We shall first note that Eq.~\eqref{eq:linDensity} admits a conserved energy $E$. It is easy to show that the energy $E$ is given by [\refcite{soliton}]:
	\begin{gather}
		E = \frac{1}{2}	\int_\reals \left[\frac{\rho_t^2}{c_s^2} + \rho_r^2\right](a^2+r^2)dr,\label{eq:energy}
	\end{gather}
hence we consider solutions of $\rho$ vanishing at boundaries i.e. $\rho(t,r)\sqrt{a^2+r^2}\rightarrow 0$ under limit $r\rightarrow\pm\infty$, and also s.t. $E$ exists, hence: $\rho(t,r)\sqrt{a^2+r^2} \in H_0^1(\reals)$, that is: the first derivatives of $\rho(t,r)\sqrt{a^2+r^2}$ are also square integrable for each $t$.

	We now move to constructing the operator similar to the one shown in [\refcite{wavemaps}]. We begin by first introducing an auxiliary function $\chi(r)$ defined to be:
	\begin{gather}
		\chi(t,r) = \rho_1(t,r)\sqrt{a^2+r^2}.\label{eq:auxiliary}
	\end{gather}
	This allows us to reduce the order of the first order term in the hyperbolic PDE Eq.~\eqref{eq:linDensity}, thus yielding an equation for $\chi(r)$:
	\begin{gather}
		\frac{1}{c_s^2}\frac{\partial^2\chi}{\partial t^2} - \frac{\partial^2\chi}{\partial r^2} + \frac{a^2}{(a^2+r^2)^2}\chi = 0. \label{eq:auxiliaryPDE}
	\end{gather}
	We consider an ansatz solution: $\chi(t,r) = e^{\lambda c_st}w(r)$, for some $\lambda\in\complex$ and $w\in H_0^1(\reals)$ reducing Eq.~\eqref{eq:auxiliaryPDE} to the following Sturm-Liouville problem:
	\begin{gather}
		 Lw = \left[-\partial_r^2 + \frac{a^2}{(a^2+r^2)^2}\right]w = -\lambda^2 w,\label{eq:sturmliouville}
	\end{gather}
where $L$ is a Sturm-Liouville (SL) operator. Let $\displaystyle q(r) = \frac{a^2}{(a^2+r^2)^2}$ which is non-negative everywhere. Then for a solution $w \in H_0^1(\reals)$ we have:
	\begin{gather}
		-\lambda^2 (w,w) = (Lw,w) = \int_\reals	\left[-w\partial_{r}^2w + q(r)w^2\right]dr = \int_\reals \left[w_r^2 + q(r)w^2\right]dr \geq 0,\nonumber
	\end{gather}
	indicating that all eigenvalues of $L$ are non-negative, hence $\lambda$ is purely imaginary and the linear system is subject to oscillations around the zeroth order solutions.

\begin{figure}[h]
	\begin{center}$
		\begin{array}{c}
		\includegraphics[width=80 mm]{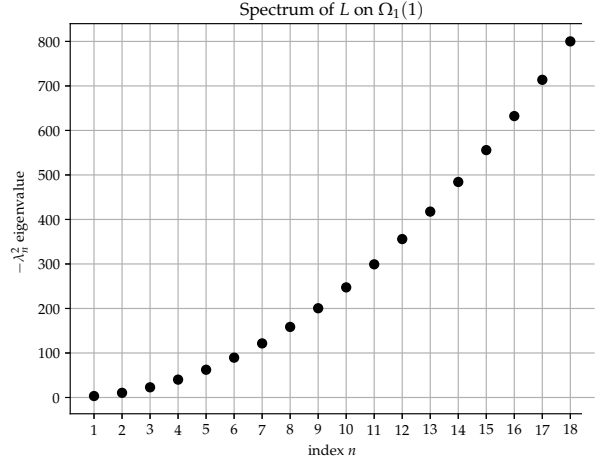}
		\end{array}$
	\end{center}
	\caption{Spectrum of Eq.~\eqref{eq:sturmliouville} on $\Omega_1(1)$.}
	\label{fig:spectrum}
	\end{figure}

Another issue we have to address is the spectral properties of Eqs.~\eqref{eq:linDensity} which is a fingerprint of the corresponding sound wave. We begin by considering a simpler, finite domain case, where the fluid is trapped inside a bounded subspace $\Omega_R(a)$ of $\w{a}$ of radius $R$, defined to be:
	\begin{gather*}
		\Omega_R(a) = \left\{(t,r)~|~ (t,r,\theta,\phi)\in\w{a},~|r| < R \right\}.
	\end{gather*}
	Hence, using the Eqs.~\eqref{eq:auxiliaryPDE}, we can define the following boundary value problem (BVP):
	\begin{gather}
		\begin{cases}
			\displaystyle\frac{1}{c_s^2}\frac{\partial^2\chi}{\partial t^2} - \frac{\partial^2\chi}{\partial r^2} + \frac{a^2}{(a^2+r^2)^2}\chi = 0, \quad \text{on}~[0,\infty)\times\Omega_R(a) \\
			\displaystyle \chi_t(0,r) = h(r),\quad \chi(0,r) = g(r),\qquad \text{where}~|r|<R \\
			\displaystyle \chi(t,\pm R) = 0,
		\end{cases}\label{eq:BVP}
	\end{gather}
given that $\overline{\Omega}_R(a)$ is compact, the Eq.~\eqref{eq:sturmliouville} becomes a regular SL problem, hence, the spectrum is discrete and there exists an increasing set of eigenvalues:
	\begin{gather*}
		0\leq -\lambda_1^2 \leq -\lambda_2^2 \leq \cdots
	\end{gather*}
	for which the solution to the regular SL problem exists. Though there is no explicit solution for the eigenvalues, they can still be found numerically.
	We do this by considering the SL problem~\eqref{eq:sturmliouville} on $\Omega_R(a)$ with $R=1$. We have numerically found the spectrum using the SLEIGN2 [\refcite{sleign2}] and Eigensystem [\refcite{eigensystem}] tools.
	The scatterplot of the eigenvalues is given on Fig.~\ref{fig:spectrum}.

\begin{figure}[h]
	\begin{center}$
		\begin{array}{c}
		\includegraphics[width=60 mm]{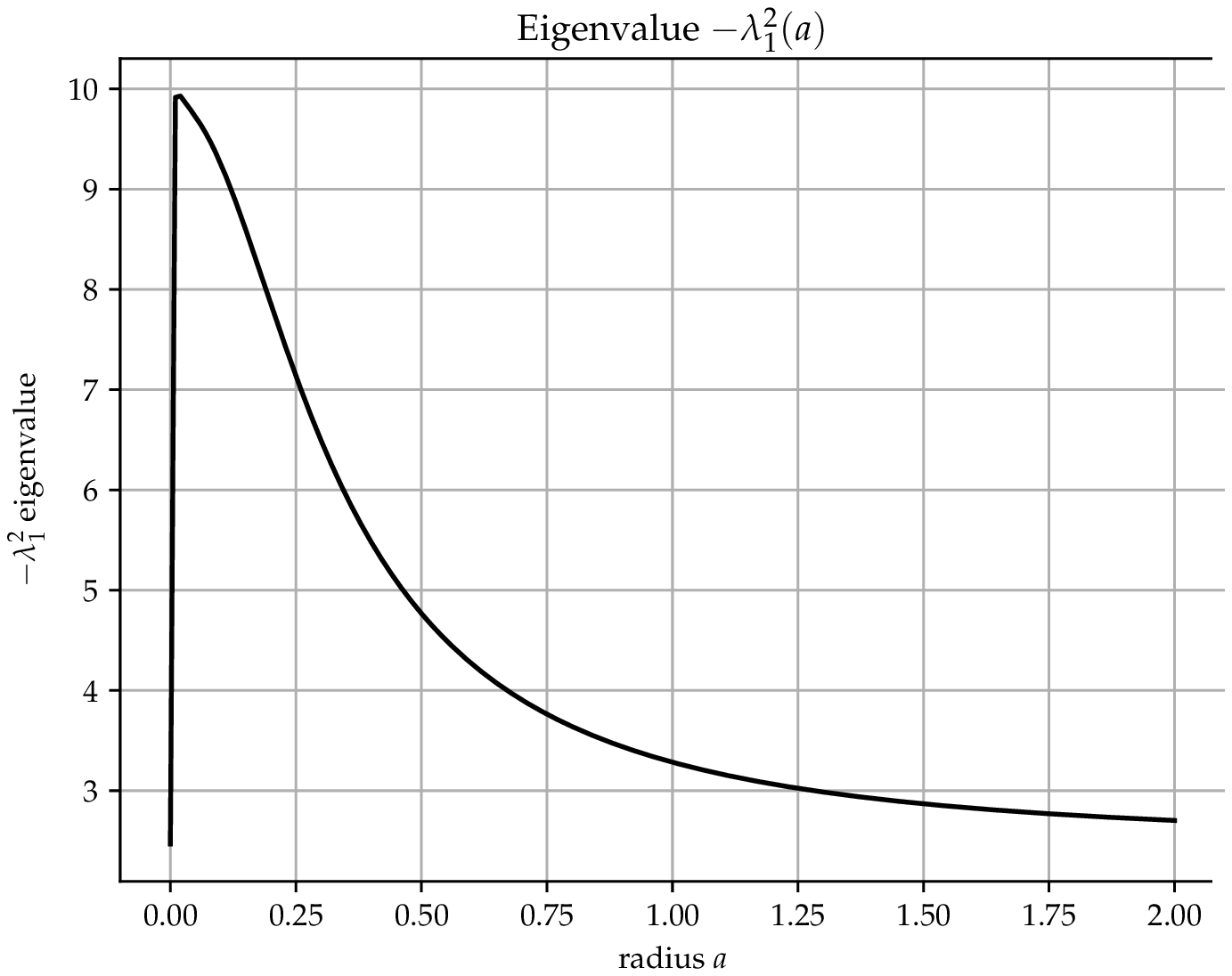}
		\includegraphics[width=60 mm]{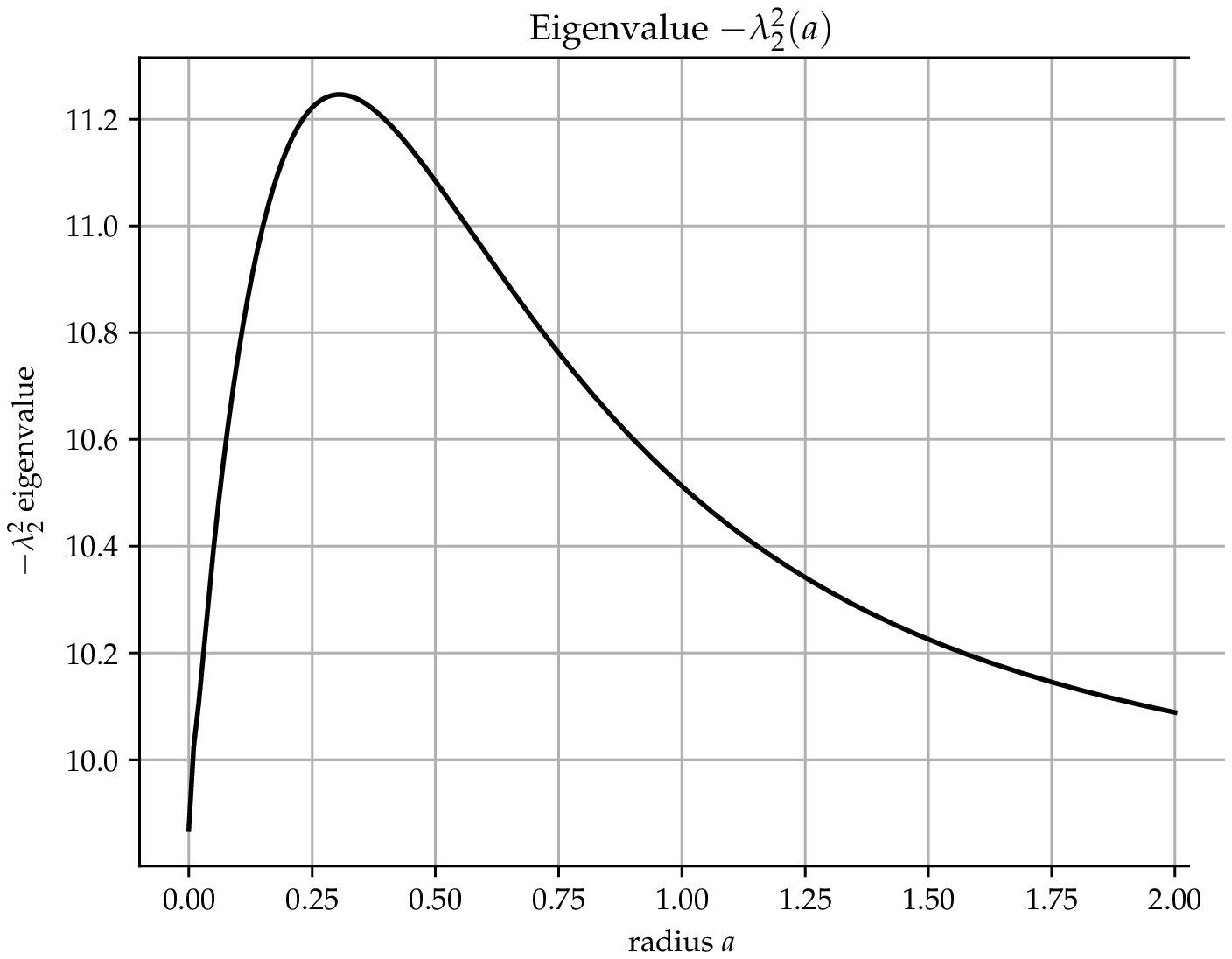} \\
		\includegraphics[width=60 mm]{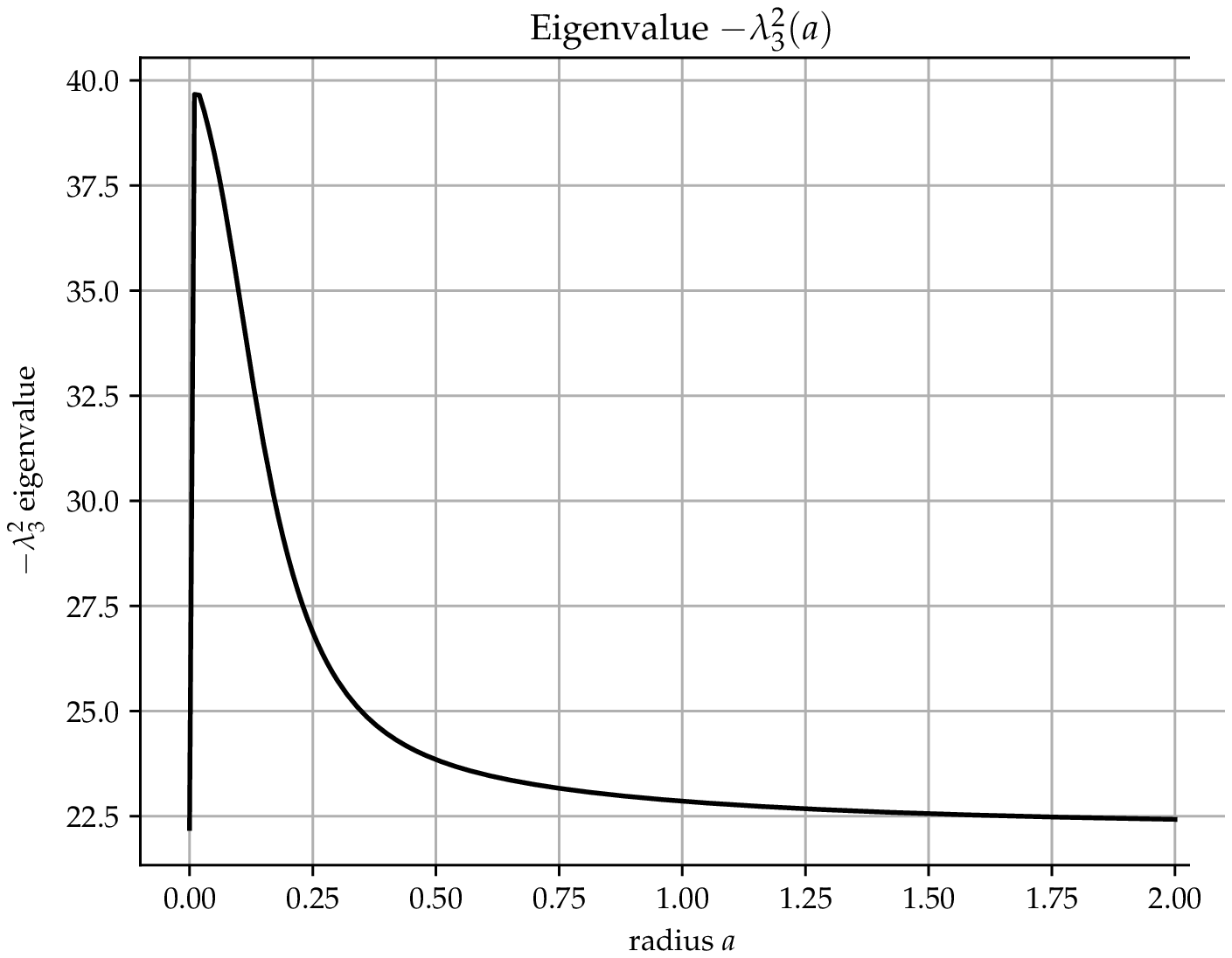}
		\includegraphics[width=60 mm]{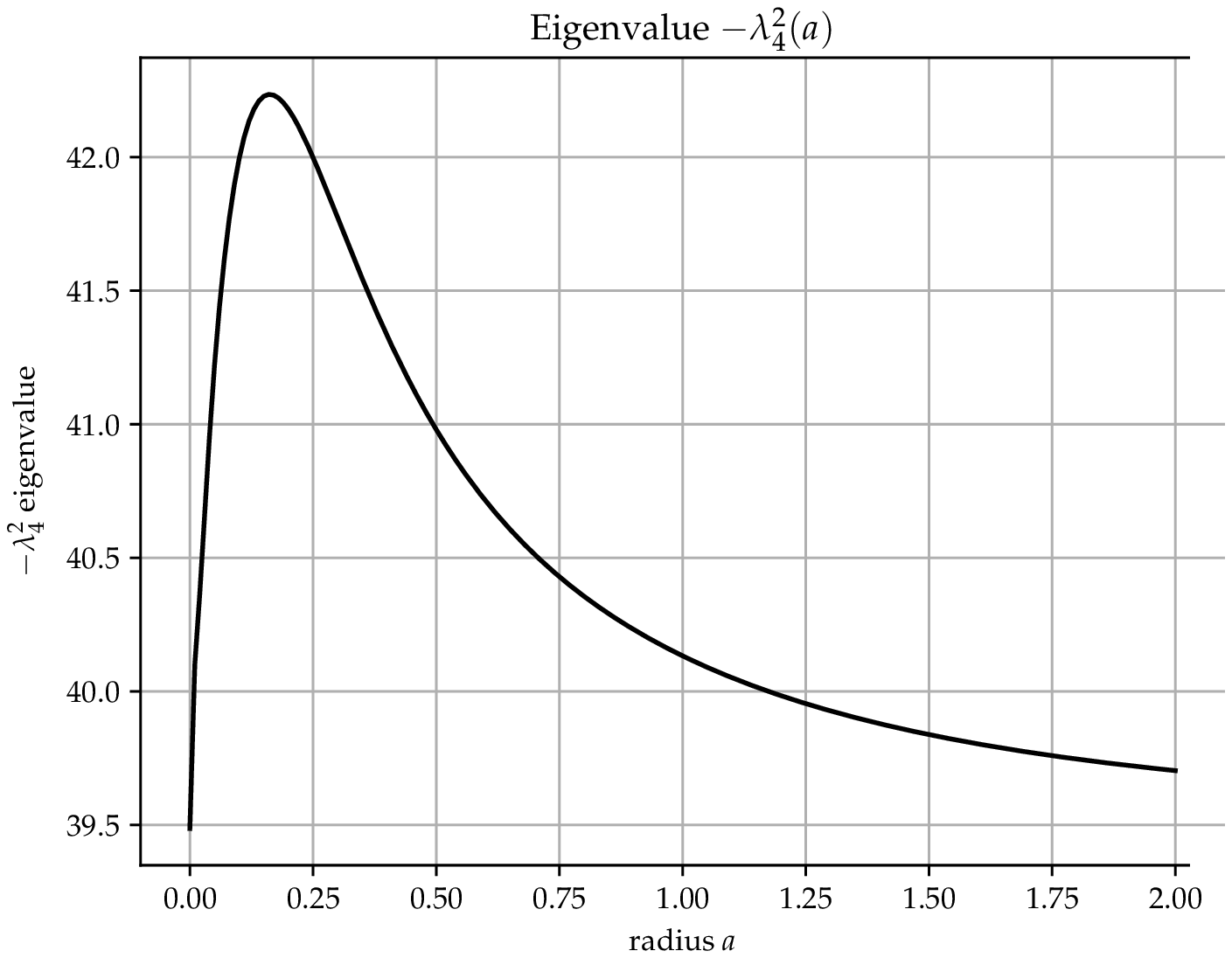}
		\end{array}$
	\end{center}
	\caption{Dependence of several $-\lambda_n^2(a)$ of Eq.~\eqref{eq:sturmliouville} on the radius $a$.}
	\label{fig:spectrum_lambda_n}
	\end{figure}

	Furthermore, we find that the eigenvalues depend on the radius of the WH in a decaying manner, which can be seen by inspecting the values of $-\lambda_n^2(a)$ at various $a$ on Fig.~\ref{fig:spectrum_lambda_n}.

\begin{figure}[h]
	\begin{center}$
		\begin{array}{c}
		\includegraphics[width=90 mm]{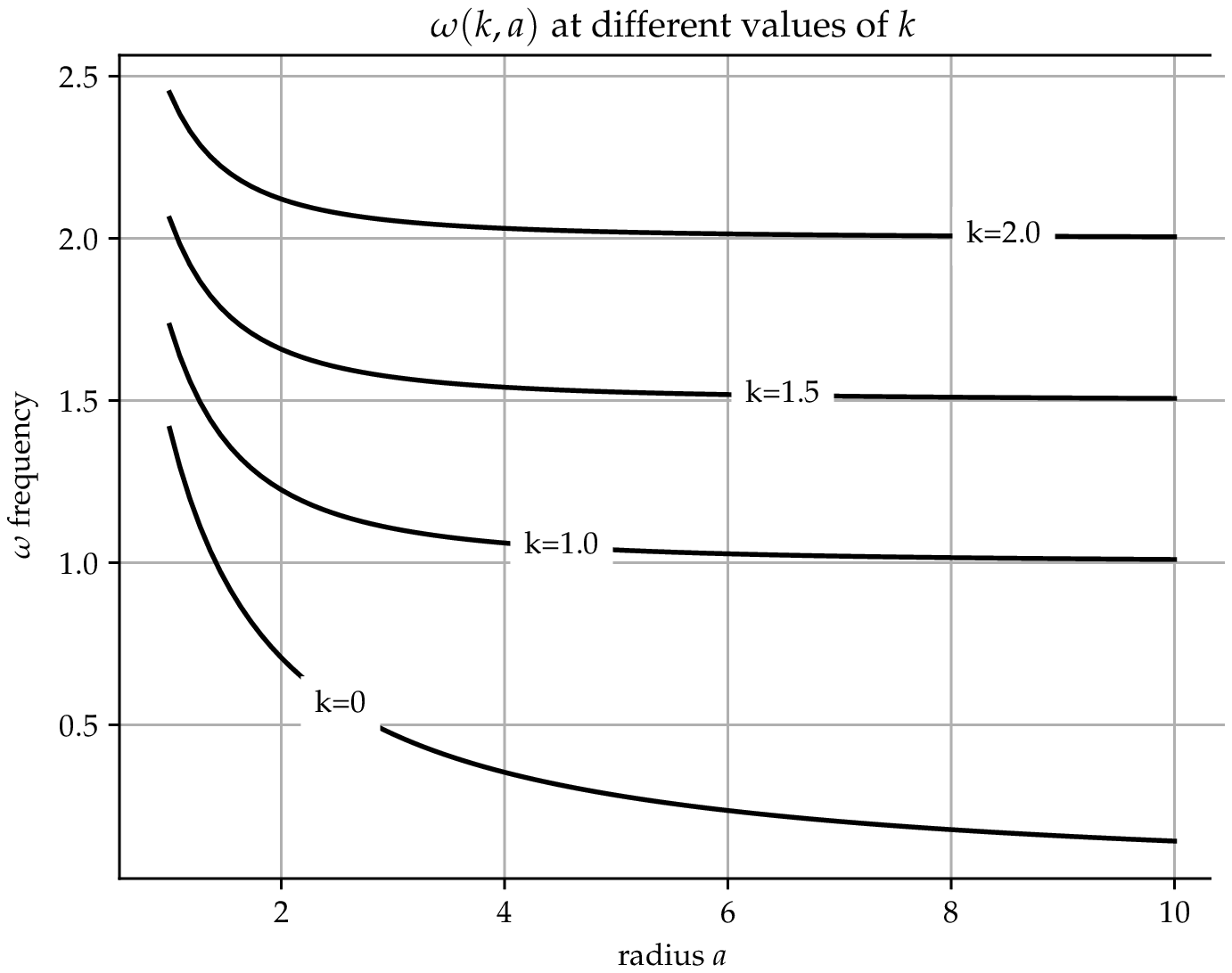}
		\end{array}$
	\end{center}
	\caption{Dependence of $\omega(k,a)$ from Eqn.~\eqref{eq:dispersion} on $a$ at different values of $k$}
	\label{fig:dispersion}
	\end{figure}

	We now move to analyzing the spectrum of Eq.~\eqref{eq:auxiliaryPDE} under limiting cases. Suppose that the fluid is again centered at the throat of the wormhole. We now consider the behaviour of the fluid in the  vicinity of the center, hence: $|r/a| \ll 1$. This allows us to expand the $q(r)$ function as:
	\begin{gather}
		q(r) = \frac{1}{a^2}\frac{1}{(1+r^2/a^2)^2} = \frac{1}{a^2} - \frac{2r^2}{a^4}+\mathcal{O}\left(\frac{r^4}{a^4}\right).\label{eq:qExpansion}
	\end{gather}
	Thus, up to the first order terms, the operator Eq.~\eqref{eq:sturmliouville} reduces to:
	\begin{gather}
		\tilde{L}^{(1)}w = \left[-\partial_r^2 + \frac{1}{a^2}\right]w = -\lambda^2w\label{eq:approxSL}.
	\end{gather}
	Therefore, under substitution $\omega^2 = -c_s^2\lambda^2$, Eq.~\eqref{eq:approxSL} yields the following dispersion relation:
	\begin{gather}
		\omega^2 = c_s^2\left(k^2 + \frac{1}{a^2}\right).\label{eq:dispersion}
	\end{gather}
	We note that, under a fixed $k$, the Eq.~\eqref{eq:dispersion} exhibits similar decaying dependence on $a$ as the eigenvalues of the general SL problem Eq.~\eqref{eq:sturmliouville} on a bounded domain. To visualize this, compare the Fig.~\ref{fig:spectrum_lambda_n} with the graph of Eq.~\eqref{eq:dispersion} on the Fig.~\ref{fig:dispersion}.

	For completeness, we shall now consider the case when $|a/r|\ll 1$, corresponding to large distances from the center of the WH. This allows us to expand $q(r)$ function as:
	\begin{gather}
		q(r) = \frac{1}{a^2}\frac{a^4}{r^4}\frac{1}{(1+ a^2/r^2)^2} = \frac{1}{a^2}\left(\frac{a}{r}\right)^4 - \frac{2}{a^2}\left(\frac{a}{r}\right)^6 + \mathcal{O}\left(\frac{a^8}{r^8}\right),\label{eq:qExpansion}
	\end{gather}
	which does not contain the first order terms, thus the operator reduces to a Helmholtz wave equation:
	\begin{gather}
		\tilde{L}w = -\partial_r^2w = -\lambda^2 w = c_s^2\omega^2 w,\label{eq:wave}
	\end{gather}
yielding no observable parameters that are intrinsic to the WH. 
Summarising, one can conclude that the sound wave's behaviour close to the central region of the WH, if observed somehow, might have characteristics influenced by the existence of the WH, which potentially can be a good sign for identifying the WH candidates.


We shall now shift our focus to analyzing the numerical solutions of the BVP first defined above in Eqs.~\eqref{eq:BVP}. Before analyzing the solutions, we first note that the SL problem given in Eq.~\eqref{eq:sturmliouville} corresponds to Time-Independent Schr\"{o}dinger's Equation with $q(r)$ potential. This motivates us to look for solutions s.t. the waves are either reflected or transmitted upon striking the center of the throat of the wormhole (i.e. $r=0$). This can be accomplished by varying the values of $a$.

\begin{figure}[h]
	\begin{center}$
		\begin{array}{c}
		\includegraphics[width=30 mm]{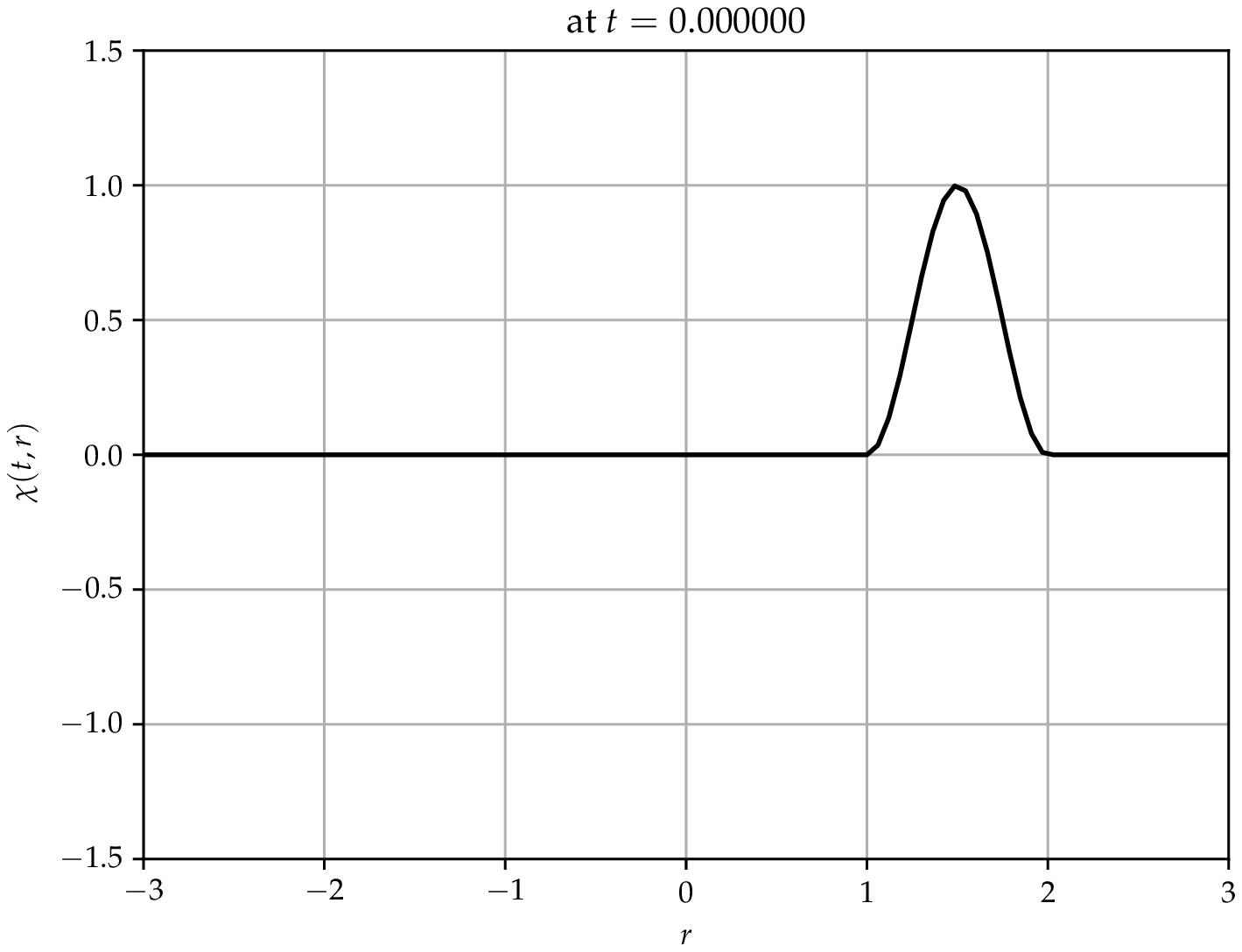}
		\includegraphics[width=30 mm]{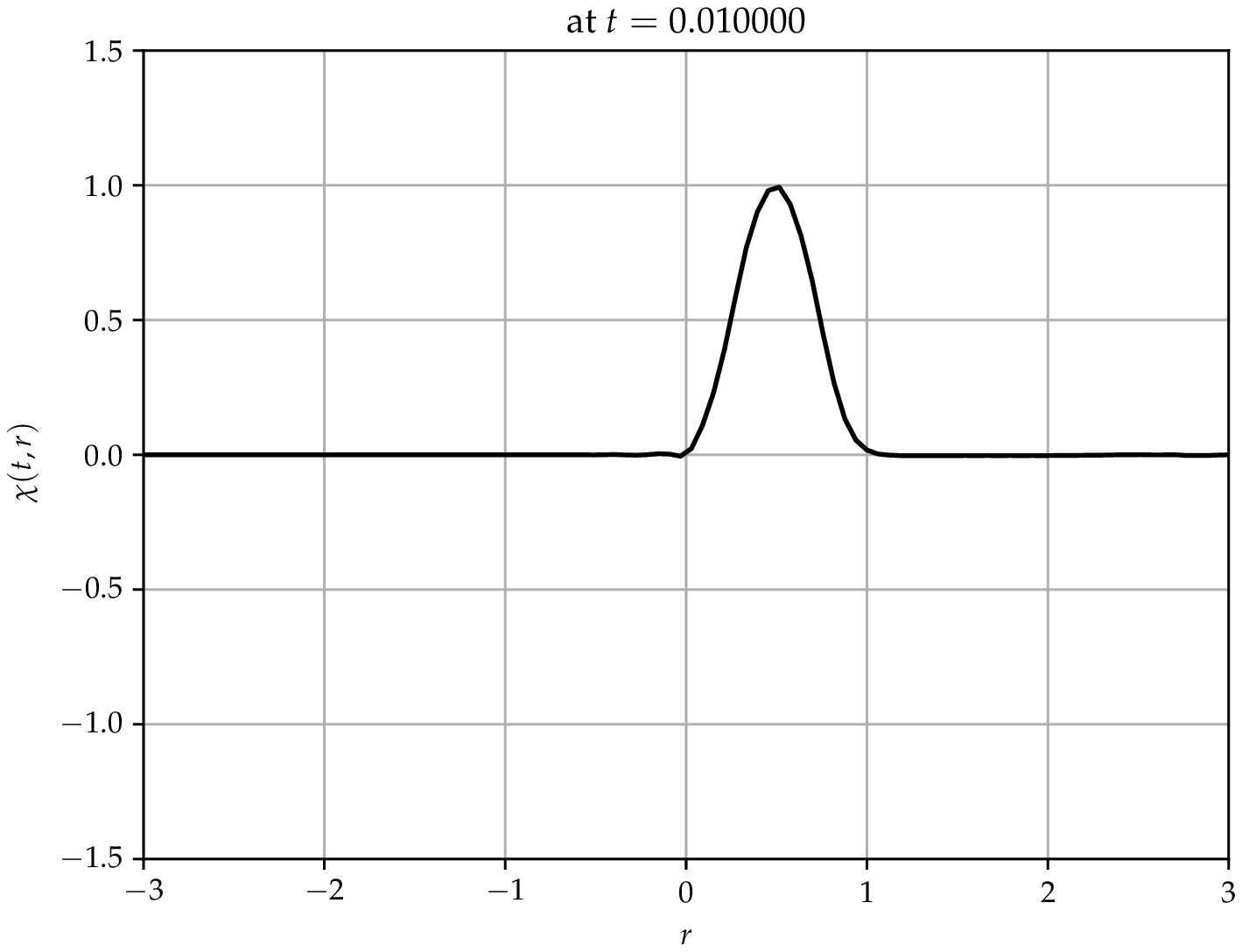}
		\includegraphics[width=30 mm]{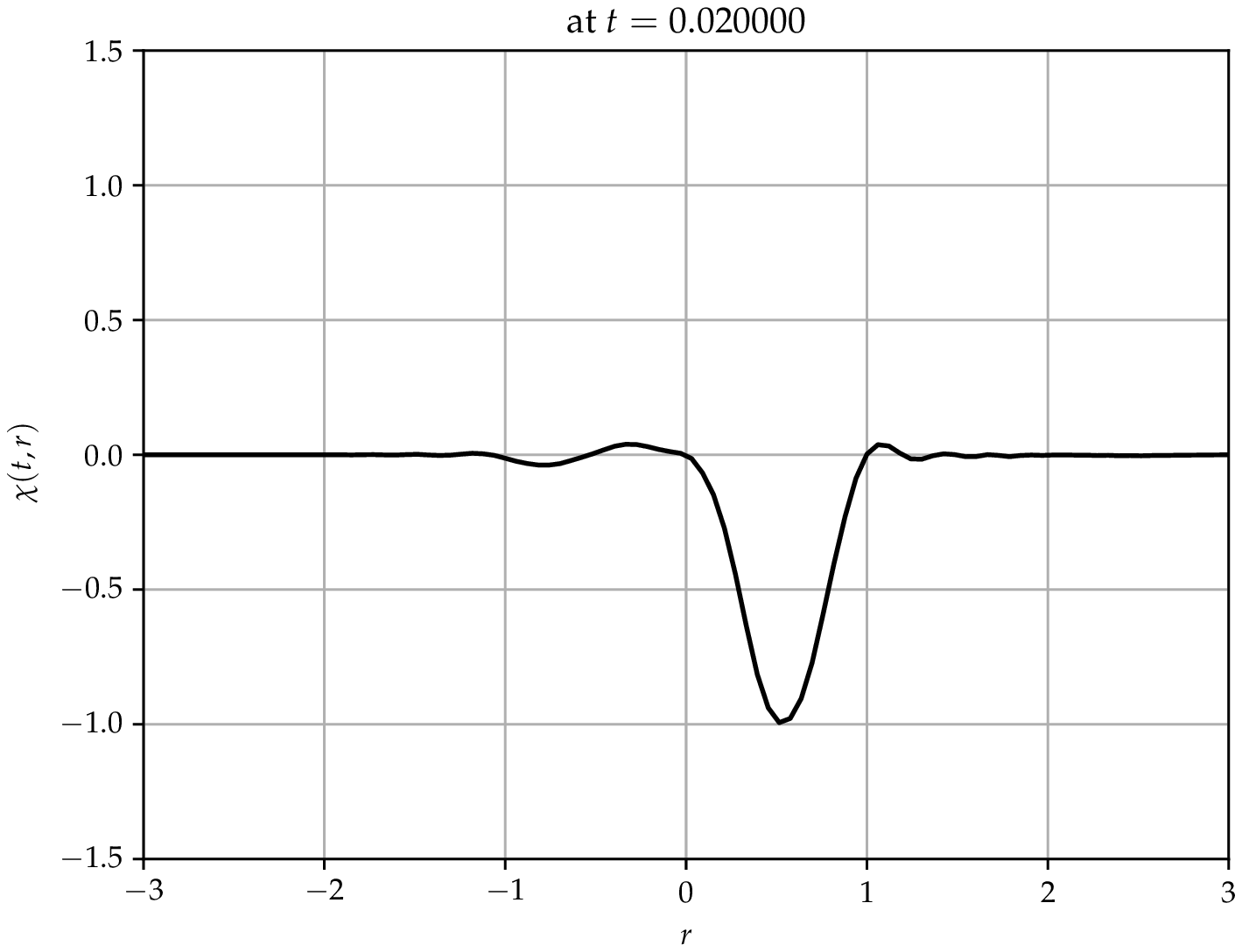}
		\includegraphics[width=30 mm]{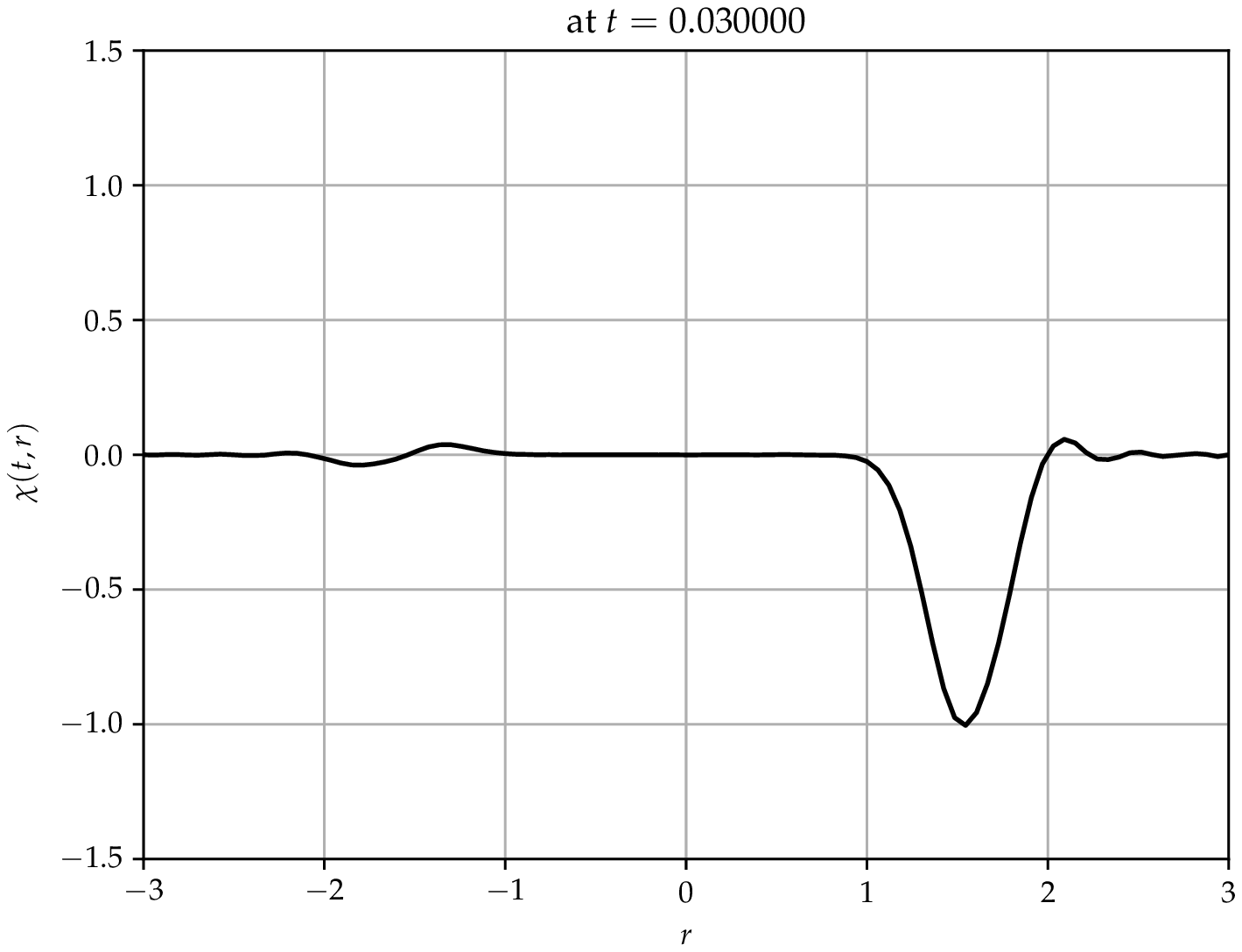}
		\end{array}$
	\end{center}
	\caption{Traveling wave-like solution of BVP.~\eqref{eq:BVP} when $a \ll 1$.}
	\label{fig:travelingWaveSmallA}
\end{figure}

To maximize the effect, we shall ensure that the solution is s.t. $-\lambda^2 < \sup_r{q(r)}$ the maximum value of potential $q(r)$, which is given by:
\begin{gather}
	\sup_{r\in\reals} q(r) = \frac{1}{a^2}
\end{gather}
Motivated with this observation, we find that a traveling wave is partially transmitted, while the rest is reflected from the origin of the WH (see Fig.~\ref{fig:travelingWaveSmallA})

\begin{figure}[h]
	\begin{center}$
		\begin{array}{c}
		\includegraphics[width=30 mm]{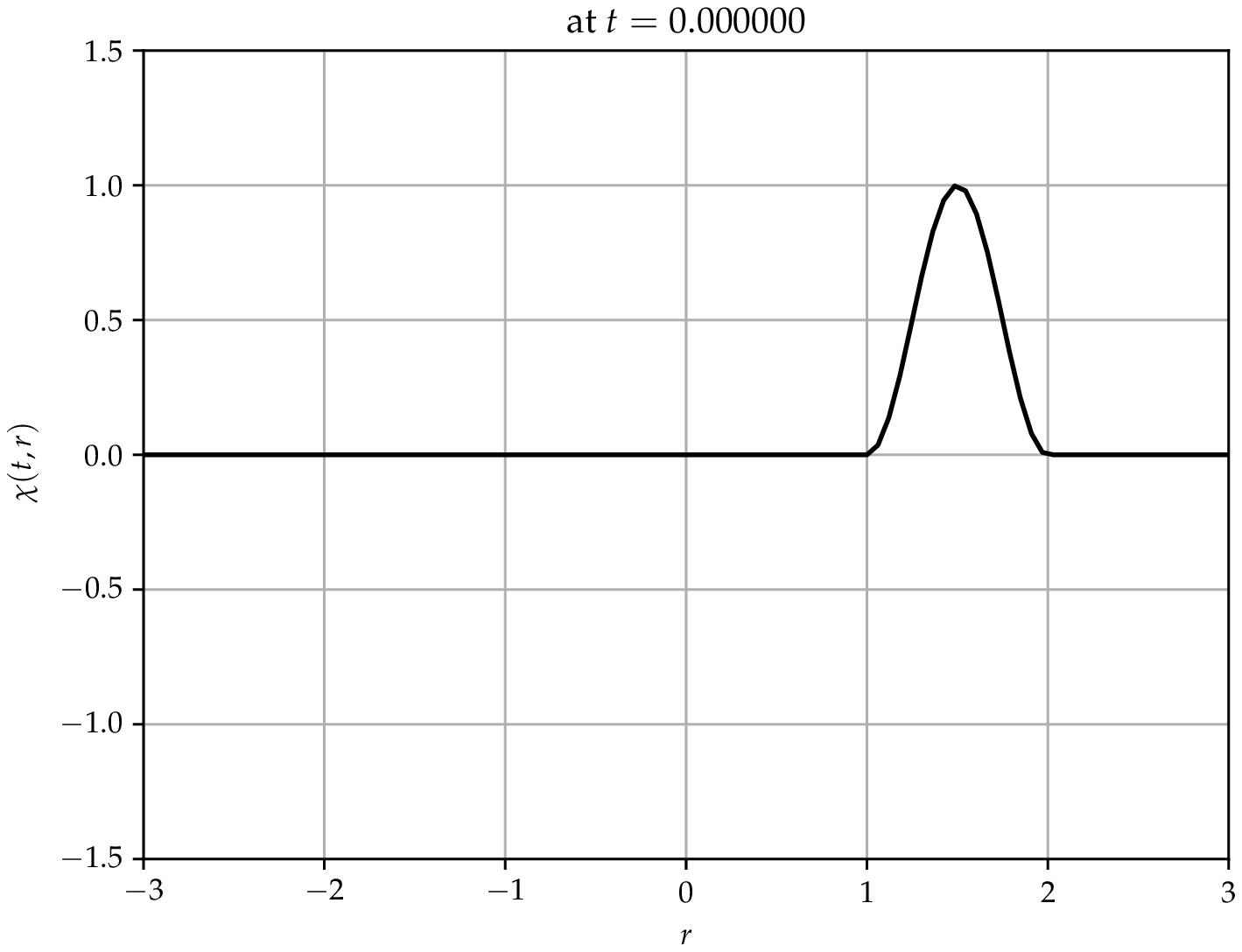}
		\includegraphics[width=30 mm]{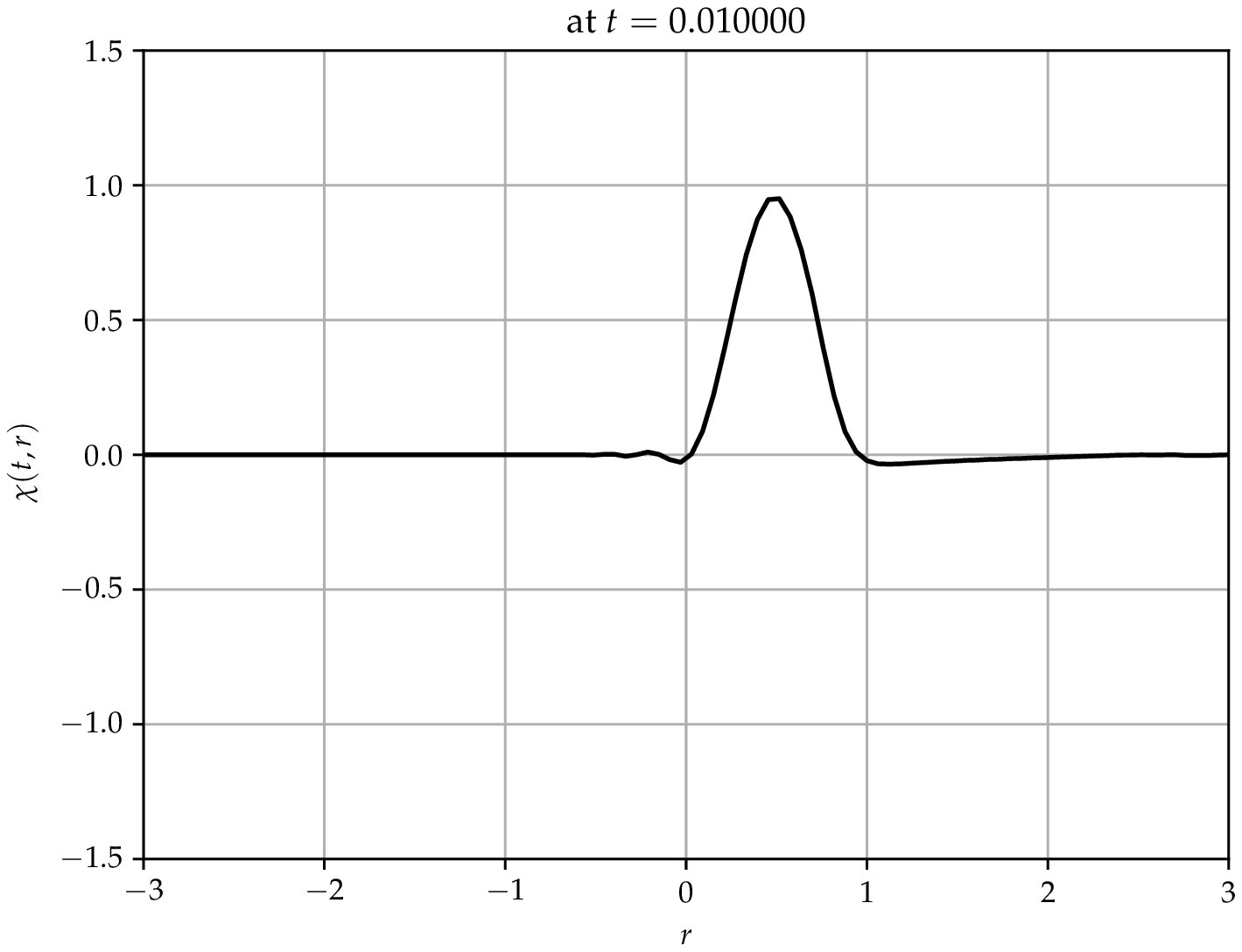}
		\includegraphics[width=30 mm]{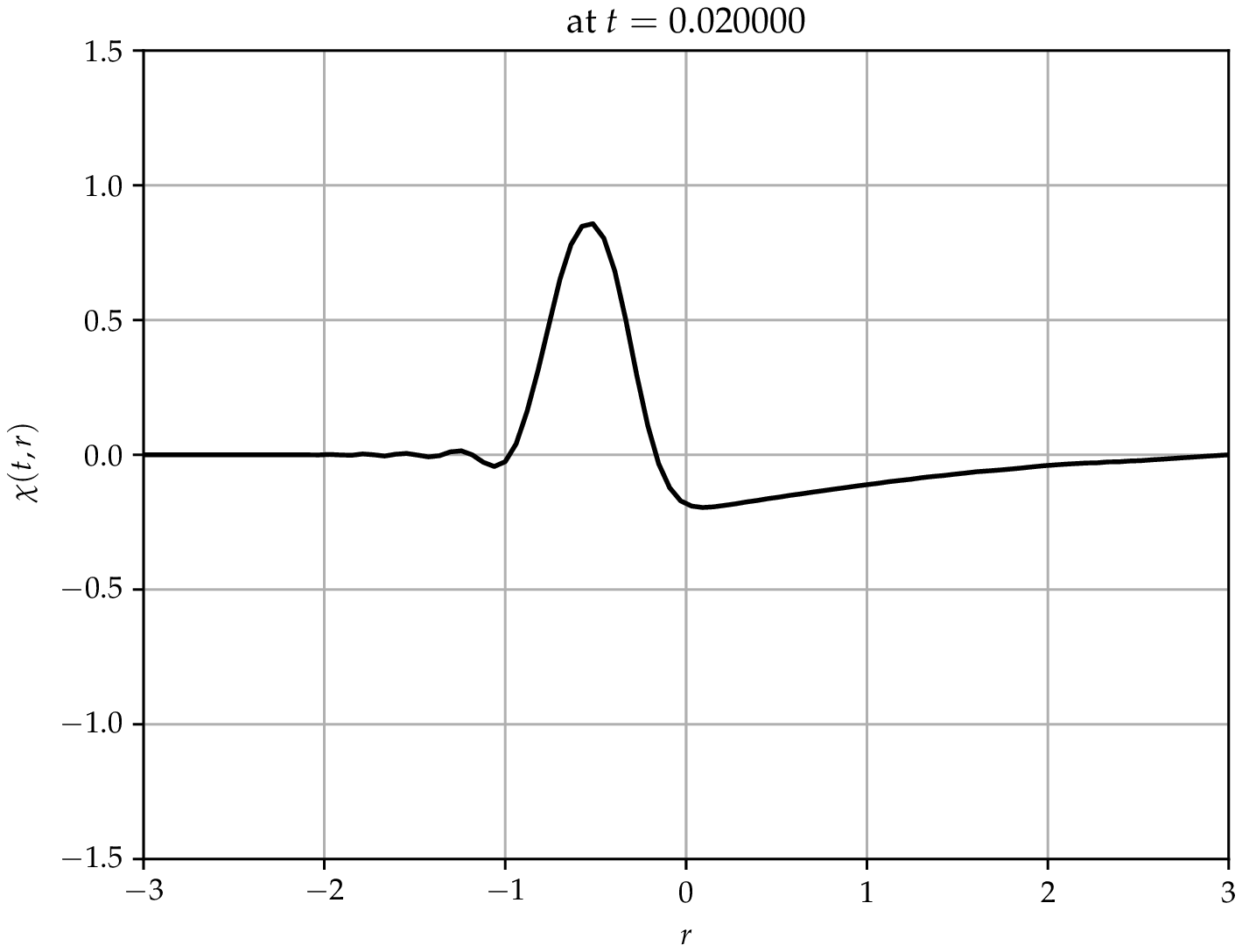}
		\includegraphics[width=30 mm]{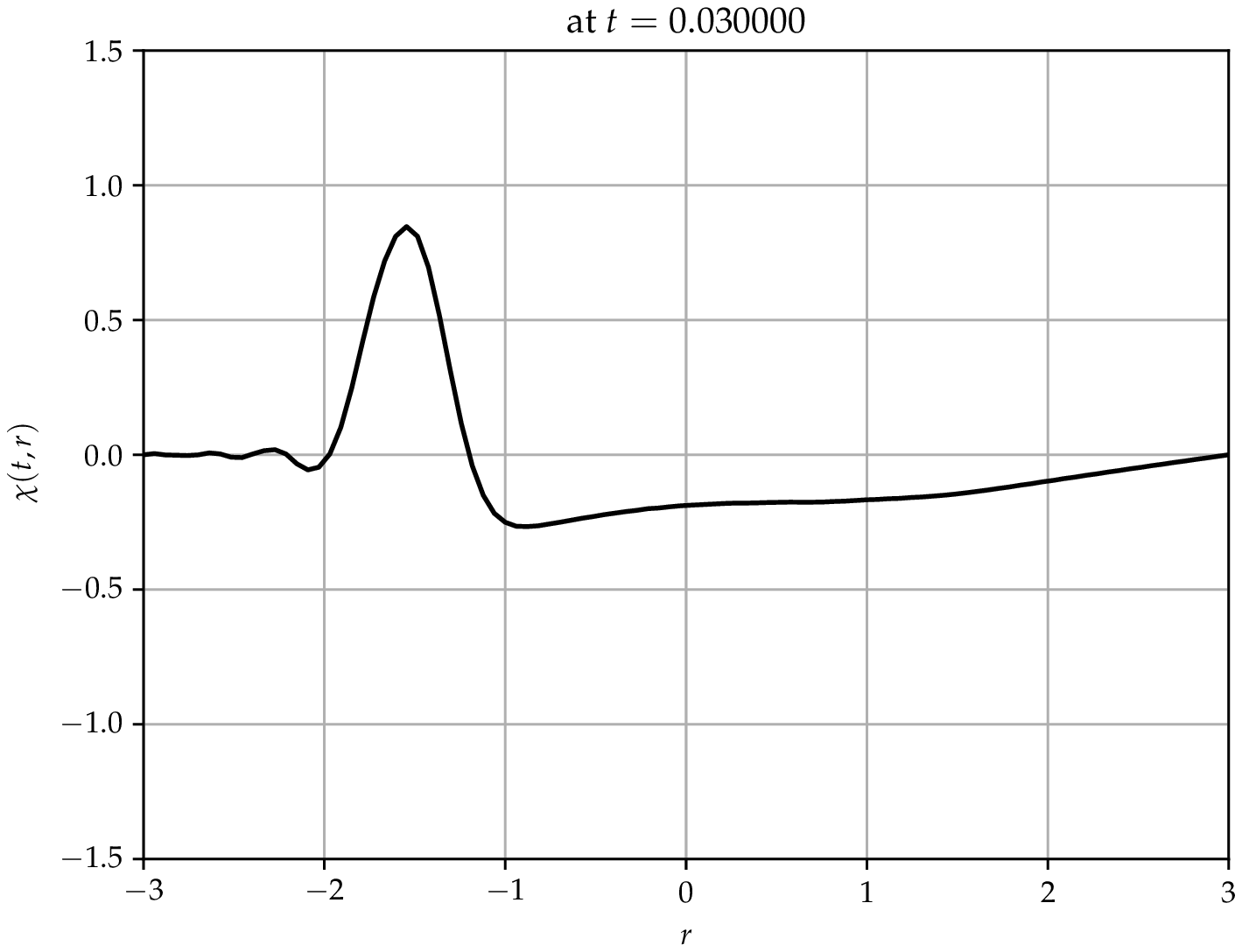}
		\end{array}$
	\end{center}
	\caption{Traveling wave-like solution of BVP.~\eqref{eq:BVP} when $a \gg 1$.}
	\label{fig:travelingWave}
\end{figure}

This effect becomes weaker as the value of $a$ grows (see Fig.~\ref{fig:travelingWave}).

Comparing Fig.~\ref{fig:travelingWaveSmallA} with Fig.~\ref{fig:travelingWave}, we see that under small values of $a$, Eq.~\eqref{eq:BVP} exhibits reflective behaviour at the origin, whereas, under large values of $a$, this effect is less prominent, however, it can still be observed.


\section{Summary}\label{sec:summary}
In this paper we have investigated the properties of an ideal inviscid fluid in Ellis-Bronnikov WH. For this purpose for relativistically hot gas we have considered the conservation laws of energy momentum and mass and examined the radial flows. 

As a first example the time-stationary solutions have been derived. We have found two types of solutions. In one of them the flow velocity is vanishing far from the center with the highest saturated density in $r\rightarrow\pm\infty$ having the minimum value at $r = 0$. Another solution corresponding to a flow with relativistic velocity and vanishing density far from the center. Both types of flows can penetrate the WH interior and leave it.

Another important result is the dispersion relation, which has been obtained for a fluid which is localized in the center of the throat of the WH.

To analyze the wave-like properties and the linear stability of the fluid, we have studied the first order perturbation of a stationary constant solution. This allowed us to construct a hyperbolic PDE and a conserved quantity - energy. 

Furthermore, we have analyzed the case when the fluid is bounded inside the wormhole by imposing a vanishing Dirichlet boundary conditions on the PDE. This allowed us to numerically find the form of the discrete spectrum of the $L$ operator and its dependence on the WH radius $a$. This dependence has been compared against the dispersion relation where a similar decaying dependence is observed for large values of $a$.

Finally, the numerical solutions of the wave equation have been analyzed. It was found that, by the analogy with the Time-independent Schr\"{o}dinger's equation, the $q(r)$ function, behaves like a repulsive potential. Moreover, under decreasing values of $a$ the center of the wormhole acts as a potential barrier which reflects incoming wave pulses.

The major aim of the paper was to consider dynamics of flows and linear waves in the most simplest WH metrics and to study only the radial dependence. It is clear that another step to extend this work is to examine the similar problems but for a general case taking into account the angular dependence of solutions. Another direction of the study should be consideration of more general WH metrics. Therefore, in the forthcoming papers we are going to examine these problems.


\section*{Acknowledgments}

The research of GB was supported by the Knowledge Foundation at Free University of Tbilisi.

\bibliographystyle{alpha-fr}

\end{document}